\documentclass[usenatbib]{mn2e}

\usepackage{epsf}
\usepackage{amsbsy}
\usepackage[dvips]{graphics}
\usepackage{amssymb}
\usepackage{amsbsy}
\usepackage{amsfonts}

\newcommand\cN{\mathcal{N}}
\newcommand\cT{\mathcal{T}}
\newcommand\p{\partial}
\newcommand\f{\frac}
\newcommand\cst{\mathrm{constant}}
\newcommand\half{{\textstyle\frac{1}{2}}}
\newcommand\third{{\textstyle\frac{1}{3}}}
\newcommand\twothirds{{\textstyle\frac{2}{3}}}
\newcommand\fourthirds{{\textstyle\frac{4}{3}}}
\newcommand\seventhirds{{\textstyle\frac{7}{3}}}
\newcommand\rmb{\mathrm{b}}
\newcommand\rmc{\mathrm{c}}
\newcommand\rmd{\mathrm{d}}
\newcommand\rmi{\mathrm{i}}
\newcommand\rmr{\mathrm{r}}
\newcommand\rms{\mathrm{s}}
\newcommand\real{\mathrm{Re}}
\newcommand\imag{\mathrm{Im}}
\newcommand\fe{\mathrm{f^e}}

\title[Viscous overstability in gaseous discs]
{Viscous overstability and eccentricity evolution in three-dimensional
gaseous discs}

\author[H. N. Latter \& G. I. Ogilvie]
{Henrik N. Latter and Gordon I. Ogilvie\\
Department of Applied Mathematics and Theoretical Physics,
University of Cambridge, Centre for Mathematical Sciences,\\
Wilberforce Road, Cambridge CB3 0WA}

\begin{document}
 
\maketitle

\label{firstpage}

\begin{abstract}
  We investigate the growth or decay rate of the fundamental mode of
  even symmetry in a viscous accretion disc.  This mode occurs in
  eccentric discs and is known to be potentially overstable.  We
  determine the vertical structure of the disc and its modes, treating
  radiative energy transport in the diffusion approximation. In the
  limit of very long radial wavelength, an
  analytical criterion for viscous overstability is obtained, which involves the
  effective shear and bulk viscosity, the adiabatic exponent and the
  opacity law of the disc. This differs from the prediction of a
  two-dimensional model. On
  shorter wavelengths (a few times the disc thickness), the criterion
  for overstability 
  is more difficult to satisfy because of the different vertical structure of
  the mode.  In a low-viscosity disc a third regime of intermediate
  wavelengths appears, in which the overstability is suppressed as the
  horizontal velocity perturbations develop significant vertical
  shear.  We suggest that this effect determines the damping rate of
  eccentricity in protoplanetary discs, for which the long-wavelength
  analysis is inapplicable and overstability is unlikely to occur on
  any scale.  In thinner accretion discs and in decretion discs around
  Be~stars overstability may occur only on the longest wavelengths,
  leading to the preferential excitation of global eccentric modes.
\end{abstract}

\begin{keywords}
  accretion, accretion discs --- hydrodynamics --- instabilities
\end{keywords}

\section{Introduction}

Accretion discs may support a wide variety of wave motions.  The basic
epicyclic tendencies of orbiting particles, which result from
gravitational and inertial forces, can combine with acoustic and
buoyancy effects to allow many different modes of oscillation.  In
some cases self-gravitational or magnetic forces may also be
significant.

Waves in accretion discs are usually analysed without regard to the
shear stress that is necessarily present in order to transport angular
momentum and facilitate accretion.  This stress, probably of turbulent
origin, is most often modelled as deriving from an effective
viscosity, which might be expected always to damp the waves.  However,
\citet{K78} found that `pulsational instability' can occur under
certain circumstances, by analogy with the `overstability' of stellar
oscillations \citep{E26}.  In particular, if the stress responds to
wavelike perturbations in a certain way, a viscous overstability is
possible, in which some of the energy being drawn by the stress from
the differential rotation of the disc is diverted into waves of
growing amplitude.  This effect has long been discussed as a means of
exciting disturbances in accretion discs and planetary rings
\citep[e.g.][]{KF80,BGT85,PS86,KHM88,PL88,ST99}.

The stresses in accretion discs and planetary rings are probably not
accurately described as arising from an isotropic viscosity in the
sense of the Navier--Stokes equation.  Nevertheless, the general
principle that overstability may occur as the stress responds to
wavelike perturbations applies to all discs, even if the details are
difficult to quantify.

Waves in thin discs are often studied in a two-dimensional
approximation in which motion in the vertical ($z$) direction,
perpendicular to the orbital plane, is absent and the disc remains in
vertical hydrostatic balance.  This approximation is highly convenient
but can only be justified under exceptional circumstances and indeed
eliminates almost all of the wave modes.  The original analysis of
\citet{K78} was in fact based on a three-dimensional disc, but at the
time of writing of his paper the behaviour of waves in such discs was
not fully understood.  Subsequently it was shown that a thin disc acts
as a waveguide that conducts in the radial direction a set of discrete
modes having distinct vertical structures determined from
an eigenvalue problem \citep{OK85,L86,LP93,KP95}.  Some of these can
be classified as f, p and g~modes by direct analogy with stellar
oscillations \citep{O98}. Generally the propagation of these waves is
restricted to certain intervals of radius because they encounter turning
points at Lindblad or other resonances.

Perhaps of greatest interest among these is the $\fe$~mode, the
fundamental mode of even symmetry about the midplane, in the
classification of \citet{O98}.  This mode is the closest equivalent of
the inertial--acoustic wave (or `density wave') that appears in the
analysis of a two-dimensional disc.  Its eigenfunction involves
horizontal velocity perturbations which are symmetric about the
midplane and have no nodes in their vertical structure, but which in general
do vary significantly with $z$ and are accompanied by vertical
velocity perturbations.  This is by far the dominant mode excited by
tidal forcing at Lindblad resonances, and its excitation, propagation
and dissipation have been investigated by \citet{LO98}, \citet{OL99}
and \citet{BOLP02}.  The $\fe$~mode is also the obvious candidate for
viscous overstability.  However, \citet{K78} assumed that the
horizontal velocities would be independent of $z$, which is not true
of the $\fe$~mode.  A self-consistent linear 
calculation of overstable modes
in a three-dimensional disc has never been carried out, though the nonlinear
behaviour of axisymmetric dics with full vertical structure has been simulated by
Kley et al. (1993). They report that when $\alpha$ is larger than a
small value the disc exhibits global,
axisymmetric, overstable modes confined to the outer edge of the
computational domain.

The question of overstability also arises in the theory of eccentric
discs.  A small eccentricity can be regarded as a special type of wave
motion, corresponding to azimuthal wavenumber $m=1$, propagating in a
circular disc.  Although the viscous overstability has been discussed
mainly for axisymmetric waves, the local dynamics of waves of modest
azimuthal mode number in a thin disc is equivalent to that of
axisymmetric waves.  Studies of eccentric viscous discs in two and
three dimensions have indeed found overstability \citep{LPP94,O01},
meaning that the eccentricity often has a negative diffusion
coefficient and would grow spontaneously and
preferentially on small radial scales (comparable to the disc
thickness).  This is contrary to the intuitive concept that viscosity
should damp eccentricity. It is important to note that $m=1$ waves
propagating in a (nearly) Keplerian disk are quite exceptional in that
they can maintain a long radial wavelength over an extended radial distance.

Understanding the growth or decay rate of
eccentricity in a disc is of great importance.  For example, in
protoplanetary systems, eccentricity is a joint property of the
planets and the disc since it is exchanged through secular and
resonant interactions.  The behaviour of eccentricity within the disc
therefore affects the eccentricity of all the planets as well.

The decretion discs around Be~stars appear to have global
eccentric modes which may be excited by the viscous overstability
\citep{K83,O91}.  However, since the overstability is believed to grow
preferentially on a wavelength comparable to the disc thickness it may
be difficult to account for the appearance of coherent global modes.

It was shown by \citet{O01} that the criterion for overstability is
different in two- and three-dimensional discs, and that it can be
suppressed either by introducing a sufficiently large bulk viscosity
or by using a more sophisticated model of the turbulent stress that
involves a sufficiently large relaxation time.  \citet{LO06} have also
shown that the viscous overstability is suppressed in dilute planetary
rings when the non-Newtonian nature of the stress is taken into
account.  The analysis of eccentric discs by \citet{O01} assumes that
the eccentricity is independent of $z$, which is correct if
sufficiently large stresses are present to couple different strata in
the disc.  In a low-viscosity disc, however, the eccentricity could
vary with $z$ and might be better described by the structure of the
$\fe$~mode.

We are led, therefore, to investigate the behaviour of wave modes in a
viscous disc.  In Section~\ref{s:2d} we analyse a simple
two-dimensional model by way of introduction, while the full
three-dimensional model is presented in Section~\ref{s:3d}.  We
describe the methods of numerical solution in
Section~\ref{s:numerical} and the results follow in
Section~\ref{s:results}.  We conclude with a summary and discussion in
Section~\ref{s:conclusion}.

\section{Two-dimensional model}
\label{s:2d}

In order to describe disturbances with wavelengths that are short
compared to the radial coordinate $r$ we use the model of the shearing
sheet \citep{GLB65}.  A point in the disc at reference radius $r_0$
and orbiting with angular velocity $\Omega_0$ is used as the origin of
a rotating Cartesian coordinate system in which $x$, $y$ and $z$
correspond to the radial, azimuthal and vertical directions.  The
local shear rate in the disc is given by Oort's first constant
$A_0=-(r_0/2)(\rmd\Omega/\rmd r)_0$.  Having selected the reference
point, we subsequently omit the subscript zero.  We assume throughout
that $\Omega>A>0$; in a Keplerian disc,
$A={\textstyle\frac{3}{4}}\Omega$.

In this section, by way of introduction and for the sake of comparison
with subsequent results, we consider a simple two-dimensional model
without self-gravity, heating or cooling.  We work with the equation
of motion,
\begin{equation}
  \Sigma(\p_tu_i+u_j\p_ju_i+2\Omega\epsilon_{i3j}u_j)=-\Sigma\p_i\Phi-\p_iP+
  \p_j\cT_{ij},
\end{equation}
and the equation of mass conservation,
\begin{equation}
  \p_t\Sigma+\p_i(\Sigma u_i)=0,
\end{equation}
where $\Sigma$, $P$, and $\cT_{ij}$ are the vertically integrated
density, pressure and viscous stress, $u_i$ is the fluid velocity, and
$\Phi=-2\Omega Ax^2$ is the tidal potential.  The viscous stress is
given by
\begin{equation}
  \cT_{ij}=\bar\nu\Sigma(\p_iu_j+\p_ju_i)+
  (\bar\nu_\rmb-\twothirds\bar\nu)\Sigma(\p_ku_k)\delta_{ij},
\end{equation}
where $\bar\nu$ and $\bar\nu_\rmb$ are the vertically averaged
kinematic viscosity and bulk viscosity.  The basic state of the disc
is given by $u_y=-2Ax$, $\Sigma=\cst$, and $\cT_{xy}=-2A\bar\nu\Sigma$.
We assume that $P$, $\bar\nu$, and $\bar\nu_\rmb$ are specified
functions of $\Sigma$, and define the sound speed $v_\rms$ through
$v_\rms^2=\rmd P/\rmd\Sigma$.

We introduce perturbations of the form $\real[\Sigma'\exp(st+\rmi
kx)]$, etc., where $s$ is the (complex) growth rate and $k$ is the
(real) radial wavenumber.  These perturbations are independent of $y$
and correspond to axisymmetric waves in cylindrical geometry, but also
accurately describe the local properties of non-axisymmetric waves of
modest azimuthal wavenumber in a thin disc.  The linearized equations
read
\begin{equation}
  \Sigma(su_x'-2\Omega u_y')=-\rmi kv_\rms^2\Sigma'+\rmi k\cT_{xx}',
\end{equation}
\begin{equation}
  \Sigma\left[su_y'+2(\Omega-A)u_x'\right]=\rmi k\cT_{xy}',
\end{equation}
\begin{equation}
  s\Sigma'+\Sigma\,\rmi ku_x'=0,
\end{equation}
\begin{equation}
  \cT_{xx}'=(\bar\nu_\rmb+\fourthirds\bar\nu)\Sigma\,\rmi ku_x',
\end{equation}
\begin{equation}
  \cT_{xy}'=-2A\f{\rmd(\bar\nu\Sigma)}{\rmd\Sigma}\Sigma'+
  \bar\nu\Sigma\,\rmi ku_y'.
\end{equation}
Solution of this algebraic system yields the dispersion relation
\begin{eqnarray}
  \lefteqn{s^3+(\bar\nu_\rmb+\seventhirds\bar\nu)k^2s^2+\left[\Omega_r^2+
  v_\rms^2k^2+\bar\nu(\bar\nu_\rmb+\fourthirds\bar\nu)k^4\right]s}&\nonumber\\
  &&+4\Omega A\f{\rmd(\bar\nu\Sigma)}{\rmd\Sigma}k^2+v_\rms^2k^2\bar\nu k^2=0,
\end{eqnarray}
where $\Omega_r^2=4\Omega(\Omega-A)$ is the square of the radial
epicyclic frequency, and is assumed to be positive.  (The symbol
$\kappa$, conventional for the epicyclic frequency, is used below to
denote opacity.)  In a Keplerian disc, $\Omega_r=\Omega$.

In the special case of an inviscid disc we obtain either a
zero-frequency mode with $s=0$ and $u_x'=0$, corresponding to a steady
`geostrophic flow' in which the Coriolis force is balanced by a
pressure gradient, or an undamped inertial--acoustic wave with
frequency $\omega$ given by the familiar dispersion relation
$\omega^2=-s^2=\Omega_r^2+v_\rms^2k^2$.

Consider first the long-wavelength limit, $k\to0$, of the dispersion
relation.  One root behaves according to
\begin{equation}
  s=-\f{4\Omega A}{\Omega_r^2}\f{\rmd(\bar\nu\Sigma)}{\rmd\Sigma}k^2+O(k^4),
\label{vi2d}
\end{equation}
which yields exponential growth (viscous instability) when
\begin{equation}
  \f{\rmd(\bar\nu\Sigma)}{\rmd\Sigma}<0,
\end{equation}
as found originally by \citet{LE74}.  The other two roots have the behaviour
\begin{eqnarray}
  \lefteqn{s=\pm\rmi\Omega_r\left(1+\f{v_\rms^2k^2}{2\Omega_r^2}\right)}&
  \nonumber\\
  &&+\f{1}{2}\left[-(\bar\nu_\rmb+\seventhirds\bar\nu)+
  \f{4\Omega A}{\Omega_r^2}\f{\rmd(\bar\nu\Sigma)}{\rmd\Sigma}\right]k^2+
  O(k^4).
\end{eqnarray}
This yields exponentially growing oscillations (viscous overstability) when
\begin{equation}
  \f{\rmd(\bar\nu\Sigma)}{\rmd\Sigma}>
  (\bar\nu_\rmb+\seventhirds\bar\nu)\f{\Omega_r^2}{4\Omega A}.
\end{equation}

Now consider shorter wavelengths.  If viscous instability occurs in
the limit $k\to0$, we reach marginal stability with $s=0$ when
\begin{equation}
  \bar\nu v_\rms^2k^2=-4\Omega A\f{\rmd(\bar\nu\Sigma)}{\rmd\Sigma}.
\end{equation}
Alternatively, if viscous overstability occurs in the limit $k\to0$,
we reach marginal stability with $s=-\rmi\omega\ne0$ when
\begin{eqnarray}
  \lefteqn{\bar\nu(\bar\nu_\rmb+\fourthirds\bar\nu)(\bar\nu_\rmb+
  \seventhirds\bar\nu)k^4+(\bar\nu_\rmb+\fourthirds\bar\nu)v_\rms^2k^2}&
  \nonumber\\
  &&=4\Omega A\f{\rmd(\bar\nu\Sigma)}{\rmd\Sigma}-
  (\bar\nu_\rmb+\seventhirds\bar\nu)\Omega_r^2.
\end{eqnarray}
In each case the right-hand side of the equation is positive while the
left-hand side is a monotonically increasing function of $k^2$,
vanishing at $k^2=0$, so we find a unique positive value of $k^2$ for
marginal stability.  The instability or overstability is quenched for
all shorter wavelengths.

For example, a Keplerian disc with constant kinematic shear
viscosity $\bar\nu\ll v_\rms^2/\Omega$ and no bulk viscosity exhibits
the viscous overstability on wavelengths $\ga9v_\rms/\Omega$, while
the maximum growth rate occurs for a wavelength
$\approx13v_\rms/\Omega$.  These wavelengths are several times the
disc thickness, partly because the case of constant $\bar\nu$ is close
to marginal stability.  We presume that the reason that the
overstability is rarely or never observed in two-dimensional numerical
simulations that include an explicit shear viscosity is either that
the domain is too small or the numerical method has an effective bulk
viscosity.  (Shearing-box simulations of turbulent discs are
invariably too small in the radial direction to detect any
overstability.)

The shearing-sheet approximation is valid only for wavenumbers
satisfying $|kr|\gg 1$.  For wavelengths comparable to $r$, not only
must the effects of cylindrical geometry be considered but also the
global structure of the disc and the radial boundary conditions.  For
our purposes, the shearing sheet correctly describes the local
properties of waves in discs.

\section{Three-dimensional model}
\label{s:3d}

Our full treatment is based on a three-dimensional shearing sheet.  We
omit self-gravity but include viscous heating and radiative cooling.

\subsection{Basic equations}

We work with the equation of motion,
\begin{equation}
  \rho(\p_tu_i+u_j\p_ju_i+2\Omega\epsilon_{i3j}u_j)=-\rho\p_i\Phi-\p_ip+
  \p_jT_{ij},
\end{equation}
the equation of mass conservation,
\begin{equation}
  \p_t\rho+\p_i(\rho u_i)=0,
\end{equation}
and the thermal energy equation,
\begin{equation}
  (\gamma-1)^{-1}(\p_tp+u_i\p_ip+\gamma p\p_iu_i)=T_{ij}\p_iu_j-\p_iF_i,
\end{equation}
where 
\begin{equation}
  T_{ij}=\mu(\p_iu_j+\p_ju_i)+(\mu_\rmb-\twothirds\mu)(\p_ku_k)\delta_{ij}
\end{equation}
is the viscous stress ($\mu$ and $\mu_\rmb$ being the dynamic shear
and bulk viscosities), $\Phi=-2\Omega Ax^2+\half\Omega_z^2z^2$ is the
tidal potential ($\Omega_z$ being the vertical epicyclic frequency,
equal to $\Omega$ in a Keplerian disc),
\begin{equation}
  F_i=-\f{16\sigma T^3}{3\kappa\rho}\p_iT
\label{f}
\end{equation}
is the radiative energy flux density, and the other symbols have their
usual meanings.  We assume an ideal gas with equation of state
$p=R\rho T$ and with $\gamma=\cst$.  The properties $\mu$, $\mu_\rmb$
and $\kappa$ are regarded as functions of $\rho$ and $T$.

\subsection{Equilibrium disc}

The basic state of the disc is given by $u_y=-2Ax$, $\rho=\rho(z)$,
$T=T(z)$, $T_{xy}=-2A\mu$, and $F_z=F_z(z)$.  For hydrostatic
equilibrium,
\begin{equation}
  \p_zp=\rho g_z=-\rho\Omega_z^2z,
\end{equation}
and for thermal equilibrium,
\begin{equation}
  \p_zF_z=4A^2\mu.
\end{equation}
These conditions are solved together with the constitutive relations
to determine the equilibrium model.  We adopt the `zero boundary
conditions' whereby $\rho=T=0$ at the surfaces $z=\pm H$ of the disc.
These are appropriate to describe an optically thick disc and avoid
the complications of matching to an atmospheric model.  The surface
density is
\begin{equation}
  \Sigma=\int_{-H}^H\rho\,\rmd z,
\end{equation}
and we define the vertically averaged kinematic viscosity $\bar\nu$
(and analogously $\bar\nu_\rmb$) via
\begin{equation}
  \bar\nu\Sigma=\int_{-H}^H\mu\,\rmd z.
\end{equation}

Under convenient assumptions the problem can be reduced to a standard
dimensionless form \citep{O01}.  Let the viscosity be given by
$\mu=\alpha p/\Omega$ and $\mu_\rmb=\alpha_\rmb p/\Omega$, where
$\alpha$ and $\alpha_\rmb$ are dimensionless constants, and let
$\kappa=C_\kappa\rho^XT^Y$, where $C_\kappa$, $X$ and $Y$ are
constants.  In particular, for Thomson (electron scattering) opacity,
$X=Y=0$.  The method of solution for the vertical structure is
described in Appendix~\ref{s:vertical}, where it is shown that
$\bar\nu\Sigma\propto\Sigma^{(10+3X-2Y)/(6+X-2Y)}$.

\subsection{Linearized equations}
\label{s:linearized}

As in Section~\ref{s:2d}, we introduce axisymmetric perturbations of
the form $\real[\rho'(z)\exp(st+\rmi kx)]$, etc., and obtain the
linearized equations
\begin{equation}
  \rho(su_x'-2\Omega u_y')=-\rmi kp'+\rmi kT_{xx}'+\p_zT_{xz}',
\label{ux'}
\end{equation}
\begin{equation}
  \rho\left[su_y'+2(\Omega-A)u_x'\right]=\rmi kT_{xy}'+\p_zT_{yz}',
\end{equation}
\begin{equation}
  \rho su_z'=\rho'g_z-\p_zp'+\rmi kT_{xz}'+\p_zT_{zz}',
\end{equation}
\begin{equation}
  s\rho'+u_z'\p_z\rho+\rho\Delta=0,
\end{equation}
\begin{equation}
  sp'+u_z'\p_z p+\gamma p\Delta=\cN,
\label{p'}
\end{equation}
with
\begin{equation}
  \Delta=\rmi ku_x'+\p_zu_z',
\end{equation}
\begin{equation} \label{NN}
  \cN=(\gamma-1)\left(-2AT_{xy}'+T_{xy}\,\rmi ku_y'-\rmi kF_x'-\p_zF_z'\right),
\end{equation}
\begin{equation}
  T_{xx}'=2\mu\,\rmi ku_x'+(\mu_\rmb-\twothirds\mu)\Delta,
\end{equation}
\begin{equation}
  T_{xy}'=-2A\mu'+\mu\,\rmi ku_y',
\end{equation}
\begin{equation}
  T_{xz}'=\mu(\rmi ku_z'+\p_zu_x'),
\end{equation}
\begin{equation}
  T_{yz}'=\mu\p_zu_y',
\end{equation}
\begin{equation}
  T_{zz}'=2\mu\p_zu_z'+(\mu_\rmb-\twothirds\mu)\Delta,
\end{equation}
\begin{equation}
  F_x'=-\f{16\sigma T^3}{3\kappa\rho}\rmi kT',
\end{equation}
\begin{eqnarray}
  \lefteqn{F_z'=-\f{16\sigma T^3}{3\kappa\rho}\bigg\{\p_zT'}&\nonumber\\
  &&\qquad+\left[-(1+X)\f{\rho'}{\rho}+(3-Y)\f{T'}{T}\right]\p_zT\bigg\},
\end{eqnarray}
\begin{equation}
  \f{T'}{T}=\f{p'}{p}-\f{\rho'}{\rho},
\end{equation}
\begin{equation} \label{mu'}
  \mu'=\f{\alpha p'}{\Omega}.
\end{equation}
Note that $\cN$ represents the effects of non-adiabatic heating and
cooling.

In general this eighth-order system of ordinary differential equations
must be solved numerically as an eigenvalue problem for the growth
rate $s$, with the wavenumber $k$ as a real parameter.  The boundary
conditions are that the solution should be regular at the surfaces
$z=\pm H$, which are singular points of the differential equations.
The numerically determined solutions $s(k)$ define the various
branches of the dispersion relation of the disc.

If equations~(\ref{ux'})--(\ref{p'}) are solved with the stress
perturbations $T_{ij}'$ and the non-adiabatic term $\cN$ set to zero,
we have a self-adjoint eigenvalue problem of the kind solved by
\citet{KP95}.  Indeed the structure of an optically thick disc with
zero boundary conditions is very similar to that of a polytropic disc.
We refer to this as the `inviscid problem'.

\subsection{Integral relation}

Using the preceding set of linearized equations and carrying
out various integrations by parts, we obtain the integral relation
\begin{eqnarray}
  \lefteqn{-s^2\int_{-H}^H\rho\left(|u_x'|^2+|u_z'|^2\right)\,\rmd z=}&
  \nonumber\\
  &&\int_{-H}^H\bigg\{\rho\Omega_r^2|u_x'|^2+\rho N^2|u_z'|^2+
  \f{1}{\gamma p}|u_z'\p_zp+\gamma p\Delta|^2\nonumber\\
  &&\quad-\Delta^*\cN+2s\mu\left(|\rmi k u_x'-\third\Delta|^2+
  \half|\rmi ku_z'+\p_zu_x'|^2\right.\nonumber\\
  &&\left.\qquad+|\p_zu_z'-\third\Delta|^2+|\third\Delta|^2\right)+
  s\mu_\rmb|\Delta|^2\nonumber\\
 &&\quad-\left(\f{\Omega}{\Omega-A}\right)\f{1}{\rho}|\rmi kT_{xy}'+
  \p_zT_{yz}'|^2-s^*\left(\f{\Omega}{\Omega-A}\right)\nonumber\\
  &&\quad\times\left(\mu k^2|u_y'|^2+\mu|\p_zu_y'|^2+
  2A\mu'\,\rmi ku_y'^*\right)\bigg\}\,\rmd z,
\end{eqnarray}
where
\begin{equation}
  N^2=g_z\left(\p_z\ln\rho-\f{1}{\gamma}\p_z\ln p\right)
\end{equation}
is the square of the Brunt--V\"ais\"al\"a frequency.  This relation is
closely related to equation~(13) in Kley et al. (1993). In the absence
of viscous and non-adiabatic effects, it yields a real
integral expression for the squared frequency $\omega^2=-s^2$ of the
mode, which has the usual variational property associated with
self-adjoint eigenvalue problems.  More generally, the imaginary part
of this equation yields (with $s=s_\rmr+\rmi s_\rmi$)
\begin{eqnarray}
  \lefteqn{s_\rmr\int_{-H}^H\rho\left(|u_x'|^2+|u_z'|^2\right)\,\rmd z}&
  \nonumber\\
  &&=\f{1}{2s_\rmi}\imag\int_{-H}^H\left[\Delta^*\cN+
  s^*\left(\f{\Omega}{\Omega-A}\right)2A\mu'\,\rmi ku_y'^*\right]\,\rmd z
  \nonumber\\
  &&-\int_{-H}^H\bigg\{\mu\left(|\rmi k u_x'-\third\Delta|^2+
  \half|\rmi ku_z'+\p_zu_x'|^2\right.\nonumber\\
  &&\left.\qquad\qquad+|\p_zu_z'-\third\Delta|^2+|\third\Delta|^2\right)+
  \half\mu_\rmb|\Delta|^2\nonumber\\
  &&\left.\qquad+\left(\f{\Omega}{\Omega-A}\right)\half\mu\left(k^2|u_y'|^2+
  |\p_zu_y'|^2\right)\right]\bigg\}\,\rmd z.
\end{eqnarray}
Within the first integral on the right-hand side, the first
term quantifies non-adiabatic effects, while the second quantifies
those of the underlying stress variation.  The second integral on the
right-hand side represents viscous dissipation.  We rewrite the
equation as
\begin{equation} \label{integral}
  s_r=\textsf{N}+\textsf{S}+\textsf{D},
\end{equation}
where \textsf{N}, \textsf{S}, and \textsf{D} represent the three
effects.

This relation is similar to equation~(3.7) of \citet{K78} but is
considerably more general because it does not assume that the mode is
only slightly perturbed from a solution of the inviscid problem.  It
shows that there are two effects that can promote growth of the wave.
One is the non-adiabatic term $\cN$, if it is appropriately correlated
with the velocity divergence $\Delta$.  (This effect is similar to the
mechanism of thermal overstability in stars.)  The other is the
viscosity perturbation $\mu'$ (which derives from the stress
perturbation $T_{xy}'$), if it is appropriately correlated with the
azimuthal velocity perturbation $u_y'$.  (This is the mechanism of
viscous overstability.)  These two potentially positive contributions
must compete with a number of negative definite terms which correspond
to the damping effect of the unperturbed viscosity acting on the shear
that is present in the mode.  It is evident from this competition that
modes with the least complicated vertical structure, such as the
$\fe$~mode, are the most likely to be overstable.

\subsection{Long-wavelength behaviour}

In Appendix~\ref{s:long} we present an asymptotic analysis of waves in
the long-wavelength limit, $k\to0$, in the presence of viscosity.  In
this limit most of the modes have a damping rate $-\real(s)$
comparable to $\alpha\Omega$.  Of those that do not, one behaves
according to equation~(\ref{vi2d}) and therefore yields the same
criterion for viscous instability as in the two-dimensional theory.
It does so because, to a first approximation, the wave does not
disturb the hydrostatic and thermal balances present in the
unperturbed disc.

The remaining long-wavelength solutions have a dispersion relation of
the form
\begin{equation}
  s=\pm\rmi\Omega_r+s_2k^2+O(k^4),
\end{equation}
where the constant $s_2$ is determined by an algebraic problem derived
in Appendix~\ref{s:long}.  These waves yield a criterion for viscous
overstability, $\real(s_2)>0$, that differs from the prediction of the
two-dimensional analysis.  The differences arise because the wave has
an (epicyclic) period comparable to the time-scales for establishing
the hydrostatic and thermal balances, and the vertical structure of
the disc therefore requires a dynamical treatment.  Vertical motion,
in the form of a `breathing motion' with $u_z'\propto z$, plays a
significant role in the dynamics of this mode.  The overstability
criterion for a Keplerian disc derived by this method agrees exactly
with that found by the method of \citet{O01}, who also noted the
importance of vertical motion and non-equilibrium.

As an example, for a Keplerian disc with constant (e.g.~Thomson)
opacity and $\alpha,\alpha_\rmb\ll1$, overstability occurs for
\begin{equation} \label{criterion}
  \alpha_\rmb/\alpha< (-33\gamma^2+138\gamma-97)/12.
\end{equation}
In contrast, the two-dimensional analysis (using the result
that $\bar\nu\Sigma\propto\Sigma^{5/3}$ for the case of constant
opacity) predicts overstability for $\alpha_\rmb/\alpha<8/3$. When
$\gamma=5/3$, the three-dimensional disc is more susceptible to
overstability, as then the right-hand side equals $31/9\approx 3.44$.

\subsection{Modelling the turbulent stress}
 
As the ordinary molecular viscosity is too small to be appreciable,
the viscous stress is presumably of turbulent origin. The simplest
`closure' model, and the one we have adopted here, is the alpha
prescription, which assumes a Newtonian stress--strain relation and an
isotropic effective viscosity. However, both assumptions are likely to
be poor approximations, because the turbulent stress (probably of
magnetohydrodynamic origin) should have a non-zero relaxation time
comparable to the orbital period, and the effective viscosity may be
considerably anisotropic. In principle we could apply a more
sophisticated closure \citep[e.g.][]{O01,O03}, but doing so may raise
more questions, and in fact obscure essential physics that is
model-independent. It hence pays to examine the simpler alpha model
disc first in detail. In this context the bulk viscosity is deployed
mainly as a surrogate for whatever stabilizing influences (e.g.~stress
relaxation) may act against overstability in reality.
 
\section{Numerical analysis}
\label{s:numerical}

We seek growing modes on general radial length-scales in a vertically
stratified gaseous disc with small viscosity. This task requires the
solution of two boundary-value problems. The first determines the
vertical structure of the disc equilibrium (Appendix~\ref{s:vertical},
equations~\ref{eq1}--\ref{eq2}), while the second determines the
vertical structure of the linear modes (Section~\ref{s:linearized},
equations~\ref{ux'}--\ref{mu'}). For simplicity and clarity we adopt a
constant (e.g.~Thomson) opacity model and assume the disc is
Keplerian; thus $X=Y=0$, $\Omega_z/\Omega=1$ and
$A/\Omega=3/4$. Furthermore, the adiabatic exponent, $\gamma$, is set
equal to either $7/5$ or $5/3$. Hence the main parameters to vary are
$\alpha$ and $\alpha_\rmb$ (the shear and bulk viscosities) and $k$
(the radial wavenumber). The regimes in which we are interested
require these to be small.

Two numerical methods were employed: the `shooting method' and
Chebyshev collocation. The former method was used to solve the
nonlinear equilibrium equations.  The latter method represents the
ODEs of the linear system on a Gauss--Lobatto grid and requires
solution of a generalized algebraic eigenvalue problem. As the
equations possess a singularity at the upper boundary, both methods
were supplemented at this point by an asymptotic analysis in order to
determine the correct limiting behaviour. A judicious choice of
dependent variables was also necessary in both cases. The midplane boundary
conditions follow from symmetry considerations. For example, even
modes require $u_z'= \rmd u_x'/\rmd z= \rmd u_y'/\rmd z=\rmd p'/\rmd
z= 0$ at $z=0$.

In addition, we set the `shooting method' upon the linearised
perturbation equations, so as to check the results of the Chebyshev
collocation method.  For modest values of $\alpha$ and $k$ the
agreement between them was good. However, for $k$ and $\alpha$ in the
parameter regime of greatest interest, the `shooting method' either
failed to converge to the overstable mode or settled on an eigenpair
numerically `contaminated' by a nearby mode.  This we blame on the
stiffness of the equations in this regime. The collocation method did
not encounter this difficulty. It, however, could produce eigenpairs which
were spurious or unconverged, and these required winnowing out. Also,
when examining very long waves for low viscosities, Chebyshev
collocation suffered from numerical error, which often compromised
accurate determination of the growth rate, these being typically
$O(\alpha\,k^2)$. That said, on scales for which interesting behaviour
occurred, this was not an issue: with twenty Gauss--Lobatto nodes the
overstable growth rate had converged to ten significant figures.

We present in the following section:
\begin{enumerate}
\item the vertical structure of the equilibrium (Fig.~1);
\item the behaviour of the overstable mode's growth rate as a function
of $k$ for different $\alpha$ and $\gamma$, when $\alpha_\rmb=0$,
alongside the components of the integral relation~(\ref{integral}) in
each case (Figs~2--4);
\item
eigenfunction structures for two illustrative parameter regimes
(Figs~5 and~6); and
\item the critical ratios of $\alpha_\rmb$ and $\alpha$ for the onset
of overstability (Figs~7 and~8).
\end{enumerate}

\section{Results}
\label{s:results}

\subsection{Equilibrium}

The characteristic structure of the equilibrium state, which we plot
in Fig.~1, differs little from a polytropic model,
except for the addition of the vertical heat flux, which is
necessarily zero at the midplane and approaches a constant on the
boundary. This constant and the semithickness of the disc serve as
the `eigenvalues' of the boundary value problem and
 take the approximate values $0.34$ and $2.4$ for the case considered ($X=Y=0$).

\begin{figure}
\begin{center}
\scalebox{.5}{\includegraphics{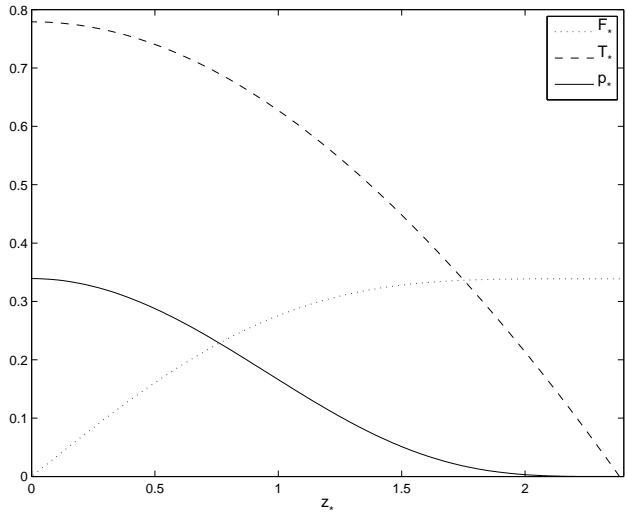}}
\caption{Non-dimensionalized pressure, $p_*$, vertical
    radiative flux, $F_*$, and temperature, $T_*$, as functions of
    non-dimensionalized height, $z_*$, for an equilibrium
    disc with constant opacity ($X=Y=0$).}
\end{center}
\end{figure}

\subsection{Growth rates and eigenfunctions}

We present the solution of the non-dimensionalized versions of
equations (\ref{ux'})--(\ref{mu'}), in which $s=s_*\Omega$,
$k=k_*/U_H$, $u'_i=u'_{i*}\Omega U_H$, while $\rho'$, $p'$, $T'$, and
$F_i'$ take the same units as their equilibrium counterparts (see
Appendix A). Since $H\approx 2.4\,U_H$, a wavenumber $k=1$
corresponds to a wavelength approximately $1.3$ times the full disc
thickness $2H$.
The growth rate is expressed in terms of the
orbital frequency, the inverse of the dynamical time-scale. The stars
will be dropped for the rest of this section.

Amongst the various even modes exhibited by the disc, our numerical
method found that only one has the potential to grow: as expected, the
viscous analogue of the inviscid $\fe$~mode, in the interval
$0<k<1$. No mode of odd symmetry grows.

In Fig.~2a we plot the real part of the growth rate of the overstable
mode, scaled by $k^2$, for modest shear viscosity and no bulk
viscosity. As $k\to 0$, this quantity approaches a constant value
approximately equal to $0.0459$, the same value predicted by the
long-wavelength theory (cf. equation~\ref{s2}).  On short enough
scales, $k\sim 1$, the mode ceases to grow. Qualitatively this
behaviour mirrors that of the two-dimensional model.

In Fig.~2b we plot the components of the integral relation
(\ref{integral}), the competition of which establishes growth or decay
on the various scales.  Immediately we note that the non-adiabatic
effects, embodied in the term \textsf{N}, are an order smaller than
those of viscous dissipation, \textsf{D}, and stress variation,
\textsf{S}. On long horizontal scales \textsf{N} contributes to
overstability, presumably through a variant of the
$\epsilon$-mechanism in stars, resulting here from the increase of
viscous dissipation in the compressed phase of the oscillation. But
on sufficiently small horizontal scales, thermal diffusion dominates
\textsf{N} and acts to smooth out disturbances (\textsf{N}$<0$).
Overstability is due almost entirely to the stress variation,
which competes with viscous damping.  This competition is most likely
to succeed on the longest wavelengths, where the mode is dominated by
horizontal motion and can avoid vertical shear.  In this regime, weak
viscous forces are sufficient to allow the mode to deviate from the
natural form of the inviscid $\fe$~mode.  On shorter wavelengths,
however, the horizontal motion necessarily develops vertical shear and
is accompanied by vertical motion, making the stress variation less
efficient in driving overstability.  The increasing horizontal shear
also acts to suppress the overstability.

\begin{figure}
\begin{center}
\scalebox{.5}{\includegraphics{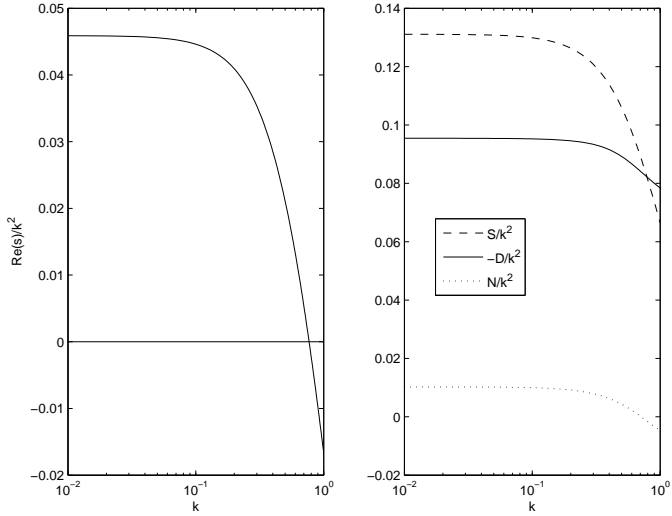}}
\caption{(a) Real part of the growth rate of the overstable mode
    divided by $k^2$, as a function of $k$, for $\alpha=0.1$,
    $\alpha_\rmb=0$, $X=Y=0$ and $\gamma=7/5$. The curve asymptotes to
    a value $\approx 0.0459$ as $k\to 0$. The growth rate $s$ is in
    units of $\Omega$, and the wavenumber $k$ is in units of
    $U_H^{-1}$ as described at the beginning of Section 5.2. 
(b) Components of the
    integral relation (\ref{integral}).  \textsf{S} refers to the
    stress variation contribution, \textsf{N} to the non-adiabatic
    contribution, and \textsf{D} to viscous dissipation.}
\end{center}
\end{figure} 

When $\alpha$ is an order smaller or less, the prevalence of
overstability on the various scales becomes more complicated, as
Figs~3 and~4 reveal.  Three
characteristic length-scales emerge, which may be divided into:
\begin{enumerate}
\item the long-wavelength limit (described in Section 3.5):
$0<k\la\alpha\ll1$;
\item an intermediate regime of anomalous viscous dissipation, and,
consequently, stability: $0<\alpha\ll k\ll 1$; and
\item a regime on shorter scales, $0<\alpha\ll k\sim 1$, in which
overstability may reappear.
\end{enumerate}

\begin{figure}
\begin{center}
\scalebox{.5}{\includegraphics{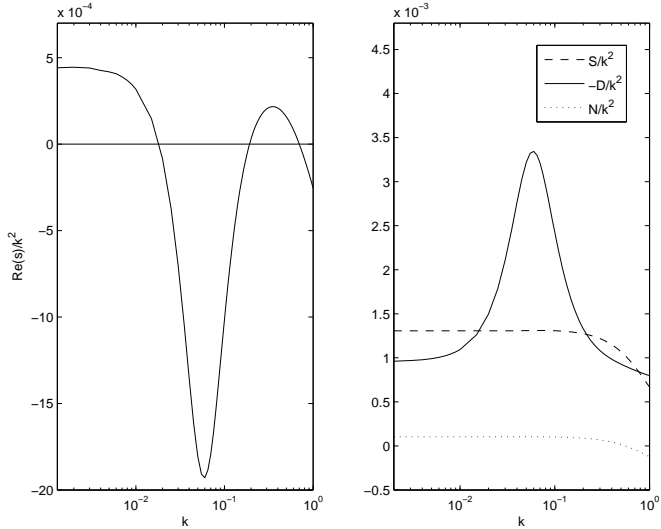}}
\caption{As for Fig.~2 but with $\alpha=0.001$, $\alpha_\rmb=0$,
    $X=Y=0$ and $\gamma=7/5$. Note the interval of intermediate
    length-scales which are stable. 
    }
\end{center}
\end{figure}

\begin{figure}
\begin{center}
\scalebox{.5}{\includegraphics{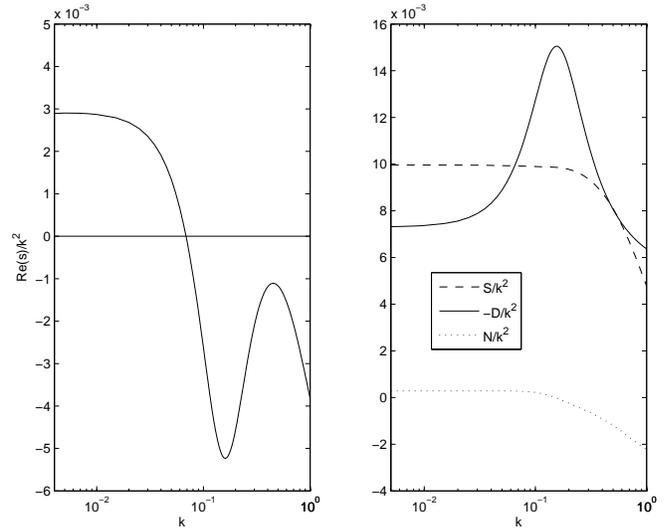}}
\caption{As for Fig.~2 but with $\alpha=0.007$, $\alpha_\rmb=0$,
    $X=Y=0$ and $\gamma=5/3$. Note that for larger $\gamma$,
    all intermediate wavelengths are stable.}
\end{center}
\end{figure}

In the long-wavelength regime, the overstable mode minimizes the
amount of vertical shear in the horizontal motion (and hence
\textsf{D}) by maintaining $u_x'$ and $u_y'$ effectively constant with
respect to $z$ (Fig.~5b). But as $k$ increases, the variation of
$u_x'$ and $u_y'$ with $z$ increases until, by $k\gtrsim\alpha^{1/2}$, they
resemble the eigenfunctions of the inviscid problem (Fig.~6). This
complication in the mode's vertical structure leads to greater viscous
dissipation and, subsequently, decay. However, once
the mode has assumed the limiting form of the
inviscid eigenfunction, the amount of shear
`saturates'.
 Consequently, the viscous dissipation plateaus, and
on shorter scales overstability may appear again.  The viscous stress in this third
regime may be considered a small perturbation to the inviscid
equations, along the lines of \citet{K78}, the principal difference
being that Kato assumed that the perturbation was with respect to a
mode with no vertical shear.

We find that the imaginary part of the growth rate is unaffected by
the increase in viscous dissipation when $\alpha$ is small. The 
long-wavelength theory correctly predicts the $O(k^2)$ correction up until
intermediate scales. This result implies that the group velocity of
the $\fe$~mode, and in particular that determining the propagation of
eccentricity in discs, is described accurately by the
three-dimensional long-wavelength theory.

Odd modes require that the horizontal components of the motion are
zero at the midplane. This means that if such modes are to minimize
their vertical shear, and hence avoid the destructive effects of
viscous dissipation, $u_x'$ and $u_y'$ must remain near zero
throughout the disc. But doing so will stall the mechanism driving
overstability, as it requires a healthy correlation between $u_y'$ and
$\mu'$. This partly explains why only even modes exhibit
overstability.

For $k\sim 1$, a larger $\gamma$ may quench overstability (Fig.~4),
whereas with long wavelengths a larger
$\gamma$ generally stimulates overstability, as equation~(\ref{criterion})
shows. The latter behaviour originates in $\gamma$'s enhancement of
the pressure perturbation in the \textsf{S} term, which, 
according to the long wavelength theory, proceeds
from its dependence on $W$ (cf. equation \ref{pres}).
For low viscosities, the adiabatic index controls
the magnitude of the vertical breathing motion via
$|W/U|=(\gamma-1)/\gamma$. Thus a larger $\gamma$ produces a larger
$W$, larger vertical pressure advection, and consequently
a larger \textsf{S}.  This effect appears more important than enhanced viscous
dissipation arising from the increase in vertical motion.

When $k\sim 1$, the mode's
response to variations in $\gamma$ may be broken into two components.
Firstly, an increase in $\gamma$ increases the scale upon which
\textsf{N} becomes negative: thermal diffusion begins to stabilize the
mode for smaller $k$. And, in fact, when $\gamma$ is sufficiently
large, the non-adiabatic term is always negative. This behaviour
springs from the sensitivity to $\gamma$ of the radiative damping
terms in $\cN$ (cf. equation~\ref{NN}).  This may arise from their
dependence on $T'$, which is larger for larger $\gamma$ (while the
other non-adiabatic contributions, and $\Delta$, do not vary
significantly).  Presumably, if the opacity is non-constant ($X$, $Y$
non-zero)
a version of the $\kappa$-mechanism of overstability in stars
may arise, though we do not explore that here.

Secondly, and more importantly, an increase in $\gamma$ may contribute
to an increase in the amount of shear in the horizontal velocities, by
analogy with the inviscid problem. This arises via $\gamma$'s
influence on the vertical stratification of the disc, and the enhanced
buoyancy forces which ensue.  \citet{LP93} established analytically
that the related `two-dimensional mode' in an inviscid isothermal disc
possesses $u_x'$ and $u_y'$ which are
$\propto\exp(N^2/\Omega_z^2)=\exp[(\gamma-1)(z/H)^2/\gamma]$.
More generally, the amount of shear in the horizontal motions of an
even inviscid p or f mode increases with $N^2$ and $z$ \citep[see
equation~13 in][]{KP95}.

\begin{figure}
\begin{center}
\scalebox{.5}{\includegraphics{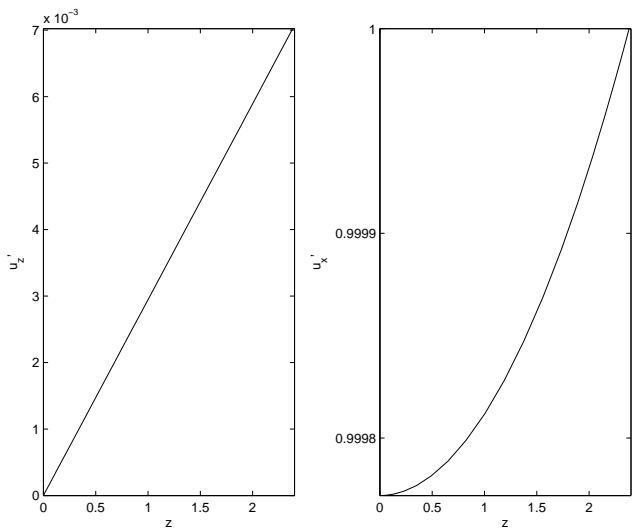}}
\caption{The $u_z'$ and $u_x'$ structure of the overstable mode for
  $\alpha=0.1$, $k=0.01$, $\gamma=7/5$ and $\alpha_\rmb=X=Y=0$. The
  eigenfunction has been normalized so that $u_x'=1$ at the
  boundary. As predicted by the long-wavelength theory, $u_z'$ is
  linear in $z$ and is $O(k)$, while the $u_x'$ eigenfunction is
  composed of an $O(1)$ constant component and an $O(k^2)$ varying
  component.  }
\end{center}
\end{figure}

\begin{figure}
\begin{center}
\scalebox{.5}{\includegraphics{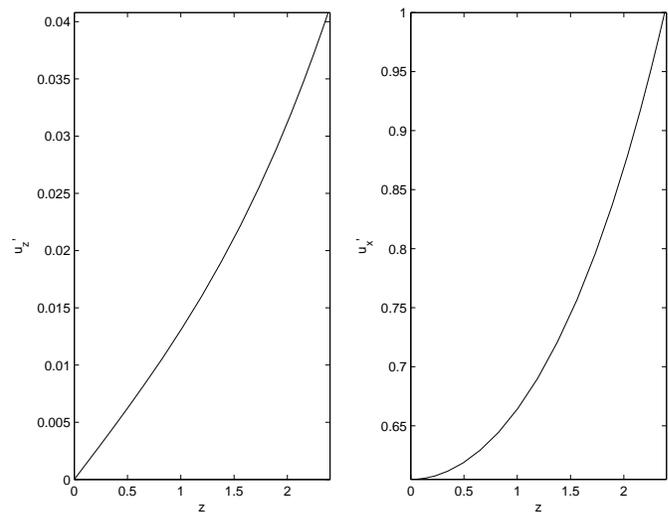}}
\caption{The $u_z'$ and $u_x'$ vertical structure for $\alpha=0.001$,
  $k=0.1$, $\gamma=7/5$ and $\alpha_\rmb=X=Y=0$. The eigenfunction
  resembles the $\fe$~mode of the inviscid problem, and decays for
  these parameters.}
\end{center}
\end{figure}

\subsection{Stability curves}

Now we let $\alpha_\rmb$ vary.
Since $\alpha_\rmb$ has a purely stabilizing effect on the
mode, a unique (but possibly negative) critical value
$(\alpha_\rmb/\alpha)_\rmc$ for marginal overstability can be
identified for given values of $\alpha$, $\gamma$ and $k$.  The
smaller the value of $(\alpha_\rmb/\alpha)_\rmc$, the more likely that
the overstability is suppressed by stabilizing influences such as
turbulent stress relaxation.  The resulting curves of marginal
stability we plot in Figs~7 and~8, for $\alpha=0.1$ and $\alpha=0.001$
respectively.

\begin{figure}
\begin{center}
\scalebox{.5}{\includegraphics{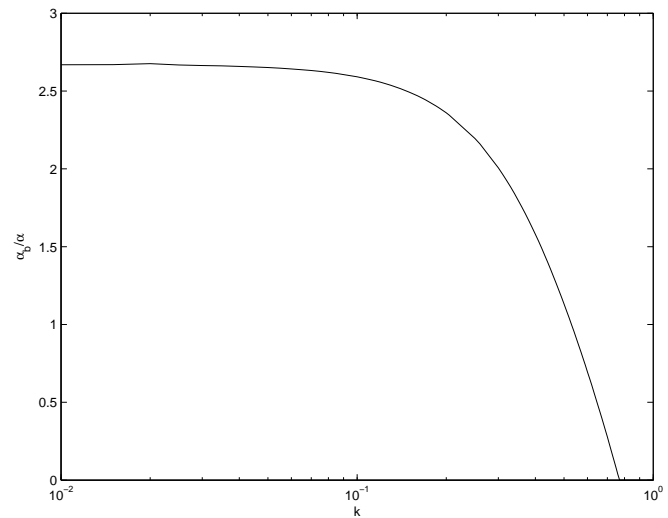}}
\caption{Curve of marginal viscous overstability in the
  $(k,\alpha_\rmb/\alpha)$ plane for $\alpha= 0.1$ and
  $\gamma=7/5$. The region above the curve is stable. As predicted by
  the long-wavelength theory, the curve approaches $197/75\approx
  2.627$ as $k\to 0$.}
\end{center}
\end{figure}

\begin{figure}
\begin{center}
\scalebox{.5}{\includegraphics{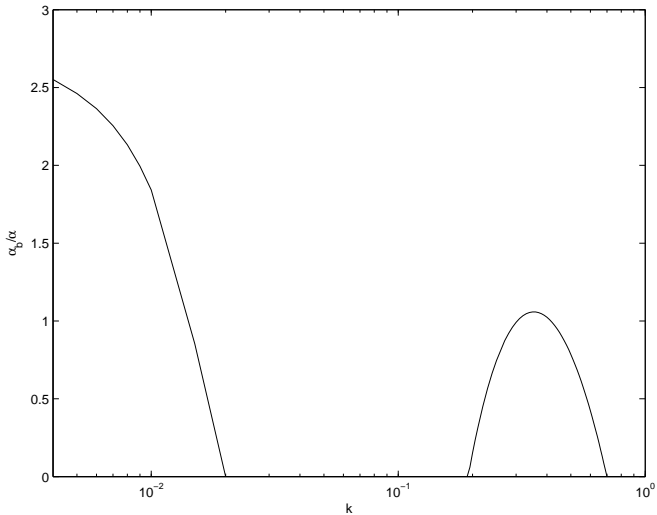}}
\caption{Curve of marginal viscous overstability in the
  $(k,\alpha_\rmb/\alpha)$ plane for $\alpha= 0.001$ and
  $\gamma=7/5$. The region above the curve is stable.}
\end{center}
\end{figure}

Both plots verify the predictions of the long-wavelength theory when
$\alpha$ is small: as $k\to 0$, $(\alpha_\rmb/\alpha)_\rmc \to 195/75$
(equation~\ref{criterion}), when $\gamma=7/5$.  Also, for
sufficiently small $\alpha$, the scales divide into the three regimes
we explored earlier (Fig.~8). We find that for small $\alpha$ and
$k\sim 1$ the
marginal curves converge to a limiting curve, differing very little
from that appearing in Fig.~8. This confirms that the stress in this
regime behaves like a regular perturbation upon the inviscid problem.

What is interesting is the relative difficulty overstability
encounters in the third regime, $k\sim 1$, when bulk viscosity is
added. An $\alpha_\rmb$ a little larger than $\alpha$ is enough to quench
overstability, whereas longer waves require a value more
than twice $\alpha$. We conclude that overstability on shorter scales
is relatively fragile: a smaller $\alpha_\rmb$ will switch it off (as
will increasing $\gamma$). This is despite the fact that unstable
modes in this regime will grow far more vigorously than those on the
longer scales.
 The simulations of Kley et al. (1993)
reflect this behaviour. The largest radial lengthscale resolvable in their
model was approximately $10H$, which falls comfortably in the
intermediate- to short-scale regime of our analysis. For sufficiently small
$\alpha$, intermediate- and short-scales will be stable for appropriate $\gamma$
and/or $\alpha_\rmb$. But, as $\alpha$ increases and we leave the
low-viscosity regime, these scales may become overstable, as Kley et
al. report.

\section{Summary and discussion}
\label{s:conclusion}

We have investigated the growth or decay rate of the fundamental mode
of even symmetry in a three-dimensional viscous accretion disc.  Our
numerical model resolves the vertical structure of the disc and its
modes, treating radiative energy transport in the diffusion
approximation.  Overstability can occur, in principle, for radial
wavelengths longer than a few times the disc thickness.  It results
almost entirely from the fact that the unperturbed disc necessarily
has a shear stress in order to facilitate accretion, and the variation
of this stress in the wave motion allows energy to be fed from the
differential rotation into growing oscillations.  This tendency
competes with the viscous damping of the mode.  Smaller contributions
to the growth rate arise from non-adiabatic effects related to the
$\epsilon$- and $\kappa$-mechanisms in stars, and thermal diffusion.

The pattern of behaviour in a low-viscosity disc ($\alpha\ll1$) is as
follows.  For radial wavenumbers $kH\sim1$ (i.e.~radial wavelengths a
few times the disc thickness) the structure of the $\fe$~mode
resembles its counterpart in an inviscid disc.  It has a non-trivial
vertical structure dictated by the stratification of the disc; the
horizontal velocities involve vertical shear and are accompanied by
vertical motion.  The mode can be overstable in this short-wavelength
regime, depending on the parameters of the disc, but the presence of
vertical shear and vertical motion acts against this tendency.

For very long radial wavelengths ($kH\la\alpha$) the mode is dominated
by horizontal motion and the vertical shear is removed by the action
of viscous stresses between different strata in the disc.  This type
of motion is optimal for exciting the overstability but the
(complicated) analytical criterion we obtain in this regime differs
from the prediction of a two-dimensional model because the disc is not
in vertical hydrostatic or thermal balance.

On intermediate wavelengths ($\alpha\ll kH\ll1$) a transitional
behaviour is found.  Here the vertical shear developing in the
horizontal velocity perturbations leads to an enhanced viscous damping
and the overstability is suppressed.

In order to bring out this subtle behaviour, we have restricted our
attention in this paper to viscous models of accretion discs based on
the Navier--Stokes equation and the alpha viscosity prescription.  We
fully appreciate that stress in accretion discs is likely to arise
from magnetohydrodynamic turbulence and to exhibit a more complicated
rheology \citep[e.g.][]{O01,O03}.  In particular, the non-zero
relaxation time of the stress is known to act against the
overstability \citep{O01} as it does in planetary rings \citep{LO06}.
In this paper we have employed a bulk viscosity as a convenient
surrogate for any such stabilizing tendencies.

A major application of this work is to the theory of eccentric
accretion discs, since a small eccentricity can be regarded as a wave
of azimuthal wavenumber $m=1$ and long radial wavelength propagating
in a circular disc. Indeed, it is only for $m=1$ waves in (nearly)
Keplerian discs that a long radial wavelength can be maintained over
an extended radial distance.
 Overstability appears as a negative diffusion
coefficient for eccentricity in the theory of \citet{O01}, which uses
approximations similar to the long-wavelength regime described above.
The approximations break down for intermediate wavelengths in a
very low-viscosity disc, because the viscous stresses are inadequate
to couple the different strata and enforce uniformity of the
eccentricity with height.

While the overstability may be suppressed by stress relaxation
effects, as suggested by \citet{O01}, an alternative possibility is
that overstability may indeed occur on very long radial wavelengths,
while shorter wavelengths require a treatment similar to that carried
out here.

In protoplanetary discs the stress is of uncertain origin, but values
in the range $10^{-4}\la\alpha\la10^{-2}$ are usually quoted based on
observed accretion rates and disc lifetimes, while the angular
semithickness is in the range $0.05\la H/r\la0.1$.  In this case the
long-wavelength analysis has no application, since the relevant
wavelengths greatly exceed the size of the disc.  Instead, the regimes
of short and intermediate wavelengths describe the behaviour of global
eccentric modes.  This means that a correct determination of the the
damping rate of eccentricity in a protoplanetary system requires a
three-dimensional treatment of the disc, allowing for vertical shear
in the horizontal velocities. We find that a typical decay rate of a
global eccentric 
mode (with a radial wavelength of $20H$) is $10^{-5}\,\Omega^{-1}$ for $\alpha=10^{-3}$ and
$2\times 10^{-6}\,\Omega^{-1}$ for $\alpha=10^{-4}$. These are only
rought estimates as we have not considered the opacity regimes
relevant to protoplanetary discs. For $\alpha=10^{-2}$ global modes
may be overstable.

In other accretion discs $H/r$ may be much smaller and $\alpha$
somewhat larger.  In this case the long-wavelength regime applies to
global eccentric modes.  The fact that the overstability is suppressed
for intermediate wavelengths means that only the largest-scale modes
may be permitted to grow.  This would neatly account for the
occurrence of global eccentric modes in decretion discs around
Be~stars.

No modes other than the $\fe$~mode were found to exhibit
overstability.  In particular, a similar mechanism would not permit
global warping of discs to be excited.

\section*{acknowledgments}

HNL would like to acknowledge the funding of the Cambridge
Commonwealth Trust.

{}

\appendix

\section{Vertical structure of the disc}
\label{s:vertical}

Dimensional analysis of the equations of vertical structure leads us
to introduce characteristic units $U$ for the variables $H$, $\rho$,
$p$, $T$ and $F$ according to
\begin{equation}
  U_H^{6+X-2Y}=\alpha R^{4-Y}\left(\f{3C_\kappa}{16\sigma}\right)
  \Sigma^{2+X}4A^2\Omega^{-1}\Omega_z^{-6+2Y},
\end{equation}
\begin{equation}
  U_\rho=\Sigma U_H^{-1},
\end{equation}
\begin{equation}
  U_p=\Omega_z^2U_H^2U_\rho,
\end{equation}
\begin{equation}
  U_T=R^{-1}\Omega_z^2U_H^2,
\end{equation}
\begin{equation}
  U_F=\alpha\, 4A^2\Omega^{-1}U_HU_p.
\end{equation}
Then let $z=z_*U_H$, $H=H_*U_H$, $\rho(z)=\rho_*(z_*)U_\rho$, etc.,
where starred quantities are dimensionless and satisfy the equations
\begin{equation} \label{eq1}
  \f{\rmd p_*}{\rmd z_*}=-\rho_*z_*,
\end{equation}
\begin{equation}
  \f{\rmd F_*}{\rmd z_*}=p_*,
\end{equation}
\begin{equation}
  \f{\rmd T_*}{\rmd z_*}=-\rho_*^{1+X}T_*^{-3+Y}F_*,
\end{equation}
\begin{equation}
  p_*=\rho_*T_*,
\end{equation}
\begin{equation} \label{eq2}
  \int_{-H_*}^{H_*}\rho_*\,\rmd z_*=1,
\end{equation}
together with the boundary conditions $\rho_*(\pm H_*)=T_*(\pm
H_*)=0$.  For reasonable values of $X$ and $Y$, this problem has a
unique solution that can be obtained numerically, with $H_*$
determined as an eigenvalue.

Near the upper surface, the limiting behaviour is of the form
$F_*\to\cst$, $T_*\propto(H_*-z_*)$,
$\rho_*\propto(H_*-z_*)^{(3-Y)/(1+X)}$,
$p_*\propto(H_*-z_*)^{(4+X-Y)/(1+X)}$.

The vertically averaged kinematic viscosity $\bar\nu$ is given by
\begin{equation}
  \bar\nu\Sigma=\int_{-H}^H\mu\,\rmd z=\f{\alpha U_pU_H}{\Omega}
  \int_{-H_*}^{H_*}p_*\,\rmd z_*,
\end{equation}
and therefore
\begin{equation}
  \bar\nu\Sigma\propto\Sigma^{(10+3X-2Y)/(6+X-2Y)}.
\end{equation}

\section{Long-wavelength limit}
\label{s:long}

We consider the limit $k\to0$ of equations~(\ref{ux'})--(\ref{p'}).
One possible scaling is
\begin{equation}
  s=k^2s_2+O(k^4),
\end{equation}
\begin{equation}
  u_x'=ku_{x1}'+O(k^3),
\end{equation}
\begin{equation}
  u_y'=ku_{y1}'+O(k^3),
\end{equation}
\begin{equation}
  u_z'=k^2u_{z2}'+O(k^4),
\end{equation}
\begin{equation}
  \rho'=\rho_0'+O(k^2),
\end{equation}
\begin{equation}
  p'=p_0'+O(k^2),
\end{equation}
which yields
\begin{equation}
  -2\Omega\rho u_{y1}'=-\rmi p_0'+\p_z(\mu\p_zu_{x1}'),
\end{equation}
\begin{equation}
  2(\Omega-A)\rho u_{x1}'=-2\rmi A\mu_0'+\p_z(\mu\p_zu_{y1}'),
\end{equation}
\begin{equation}
  0=\rho_0'g_z-\p_zp_0',
\end{equation}
\begin{equation}
  s_2\rho_0'+u_{z2}'\p_z\rho+\rho(\rmi u_{x1}'+\p_zu_{z2}')=0,
\end{equation}
\begin{equation}
  0=(\gamma-1)(4A^2\mu_0'-\p_zF_{z0}').
\end{equation}
Therefore
\begin{equation}
  s_2\int_{-H}^H\rho_0'\,\rmd z=-\rmi\int_{-H}^H\rho u_{x1}'\,\rmd z,
\end{equation}
\begin{equation}
  2(\Omega-A)\int_{-H}^H\rho u_{x1}'\,\rmd z=-2\rmi A\int_{-H}^H\mu_0'\,\rmd z.
\end{equation}
Since the hydrostatic and thermal balances are maintained in the
perturbed equations to leading order,
\begin{equation}
  \int_{-H}^H\mu_0'\,\rmd z=\f{\p(\bar\nu\Sigma)}{\p\Sigma}\Sigma_0',
\end{equation}
and so
\begin{equation}
  s_2\Sigma_0'=-\left(\f{A}{\Omega-A}\right)\f{\p(\bar\nu\Sigma)}
  {\p\Sigma}\Sigma_0'.
\end{equation}
The dispersion relation for this mode is therefore
\begin{equation}
  s=-\f{4\Omega A}{\Omega_r^2}\f{\p(\bar\nu\Sigma)}{\p\Sigma}k^2+O(k^4).
\end{equation}

Another possible scaling is
\begin{equation}
  s=s_0+k^2s_2+O(k^4),
\end{equation}
\begin{equation}
  u_x'=u_{x0}'+k^2u_{x2}'+O(k^4),
\end{equation}
\begin{equation}
  u_y'=u_{y0}'+k^2u_{y2}'+O(k^4),
\end{equation}
\begin{equation}
  u_z'=ku_{z1}'+O(k^3),
\end{equation}
\begin{equation}
  \rho'=k\rho_1'+O(k^3),
\end{equation}
\begin{equation}
  p'=kp_1'+O(k^3).
\end{equation}
The horizontal components of the equation of motion at leading order yield
\begin{equation}
  \rho(s_0u_{x0}'-2\Omega u_{y0}')=\p_z(\mu\p_zu_{x0}'),
\end{equation}
\begin{equation}
  \rho\left[s_0u_{y0}'+2(\Omega-A)u_{x0}'\right]=\p_z(\mu\p_zu_{y0}').
\end{equation}
If $u_{x0}'$, $u_{y0}'$ depend on $z$, we obtain viscously damped
epicyclic motions at this order.  The modes satisfy a Sturm--Liouville
equation and only one mode is undamped at leading order, being
independent of $z$.  For this solution, the right-hand sides vanish
and
\begin{equation}
  s_0^2+4\Omega(\Omega-A)=0.
\end{equation}
Therefore $s_0=\pm\rmi\Omega_r$, $u_{x0}'=U$ (independent of $z$) and
$u_{y0}'=V=(s_0/2\Omega)U$.  The remaining equations at leading order
yield
\begin{eqnarray}
  \lefteqn{\rho s_0u_{z1}'=\rho_1'g_z-\p_zp_1'}&\nonumber\\
  &&+\p_z\left[2\mu\p_zu_{z1}'+(\mu_\rmb-\twothirds\mu)
  (\rmi U+\p_zu_{z1}')\right],
\end{eqnarray}
\begin{equation}
  s_0\rho_1'+u_{z1}'\p_z\rho+\rho(\rmi U+\p_zu_{z1}')=0,
\end{equation}
\begin{eqnarray}
  \lefteqn{s_0p_1'+u_{z1}'\p_zp+\gamma p(\rmi U+\p_zu_{z1}')}&\nonumber\\
  &&=(\gamma-1)(4A^2\mu_1'-4A\mu\,\rmi V-\p_zF_{z1}'),
\end{eqnarray}
\begin{eqnarray}
  \lefteqn{F_{z1}'=-\f{16\sigma T^3}{3\kappa\rho}\bigg\{\p_zT_1'}&\nonumber\\
  &&\qquad+\left[-(1+X)\f{\rho_1'}{\rho}+(3-Y)\f{T_1'}{T}\right]\p_zT\bigg\},
\end{eqnarray}
\begin{equation}
  \f{T_1'}{T}=\f{p_1'}{p}-\f{\rho_1'}{\rho}.
\end{equation}
\begin{equation}
  \mu_1'=\f{\alpha p_1'}{\rho}.
\end{equation}
These can be solved, in principle, for $u_{z1}'$, $\rho_1'$, $p_1'$,
$T_1'$, $F_{z1}'$ (see below).  Then the horizontal components of the
equation of motion at order $k^2$ yield
\begin{eqnarray}
  \lefteqn{\rho(s_2U+s_0u_{x2}'-2\Omega u_{y2}')=-\rmi p_1'}&\nonumber\\
  &&+\rmi\left[2\mu\,\rmi U+(\mu_\rmb-\twothirds\mu)
  (\rmi U+\p_zu_{z1}')\right]\nonumber\\
  &&+\p_z\left[\mu(\rmi u_{z1}'+\p_zu_{x2}')\right],
\label{ux2'}
\end{eqnarray}
\begin{eqnarray}
  \lefteqn{\rho\left[s_2V+s_0u_{y2}'+2(\Omega-A)u_{x2}'\right]=-2\rmi A\mu_1'-
  \mu V}&\nonumber\\
  &&+\p_z(\mu\p_zu_{y2}').
\label{uy2'}
\end{eqnarray}
We eliminate $u_{x2}'$ and $u_{y2}'$, obtaining the solvability
condition for these equations, by taking $s_0$ times
equation~(\ref{ux2'}) plus $2\Omega$ times equation~(\ref{uy2'}), then
integrating vertically:
\begin{eqnarray}
  \lefteqn{2\Sigma s_0s_2U=\int_{-H}^H\bigg\{s_0\left[-\rmi p_1'-
  (\mu_\rmb+\seventhirds\mu)U\right.}&\nonumber\\
  &&\left.+(\mu_\rmb-\twothirds\mu)\rmi\p_zu_{z1}'\right]-
  4\Omega A\,\rmi\mu_1'\bigg\}\,\rmd z.
\end{eqnarray}
Note the contribution to this equation from the vertical motion, which
also changes the relation between $p_1'$ and $U$.

Under convenient assumptions ($R$, $\gamma$, $\alpha$, $\alpha_\rmb$
being independent of $z$) we have the analytical solution
\citep[cf.][]{O01}
\begin{equation}
  u_{z1}'=s_0Wz,
\end{equation}
\begin{equation}
  \rho_1'=-Wz\p_z\rho+A_\rho\rho,
\end{equation}
\begin{equation} \label{pres}
  p_1'=-Wz\p_zp+A_pp,
\end{equation}
\begin{equation}
  T_1'=-Wz\p_zT+A_TT,
\end{equation}
\begin{equation}
  F_{z1}'=-Wz\p_zF_z+A_FF_z.
\end{equation}
In this solution, the Lagrangian perturbations of all quantities are
proportional to the unperturbed quantities, meaning that the disc
undergoes a kind of homogeneous (but dynamical) expansion.  We find
\begin{eqnarray} \label{long1}
  \lefteqn{s_0^2W=\Omega_z^2(-2W-A_\rho+A_p)-
  (\alpha_\rmb+\fourthirds\alpha)\Omega_z^2W(s_0/\Omega)}&\nonumber\\
  &&-\rmi(\alpha_\rmb-\twothirds\alpha)\Omega_z^2(U/\Omega),
\end{eqnarray}
\begin{equation}
  s_0A_\rho=-\rmi U-s_0W,
\end{equation}
\begin{eqnarray}
  \lefteqn{s_0A_p+\gamma(\rmi U+s_0W)}&\nonumber\\
  &&=(\gamma-1)\left[4A^2(\alpha/\Omega)(A_p+W-A_F)\right.\nonumber\\
  &&\left.\qquad\qquad\quad-4A(\alpha/\Omega)\rmi V\right],
\end{eqnarray}
\begin{equation} \label{long2}
  A_F=-(1+X)A_\rho+(4-Y)(A_p-A_\rho)-W.
\end{equation}
We solve for $W$, then return to find $s_2$ using
\begin{equation}
  \int_{-H}^Hp_1'\,\rmd z=(W+A_p)P,\qquad P=\int_{-H}^Hp\,\rmd z.
\end{equation}
Therefore
\begin{eqnarray} \label{s2}
  \lefteqn{2(\Sigma/P)s_2U=-\rmi\left(1+\f{4A\alpha}{s_0}\right)(W+A_p)}&
  \nonumber\\
  &&-(\alpha_\rmb+\seventhirds\alpha)(U/\Omega)+
  (\alpha_\rmb-\twothirds\alpha)\rmi(s_0/\Omega)W.
\end{eqnarray}
From the above equations $s_2$ can be determined algebraically.

\label{lastpage}

\end{document}